\documentclass[prl,superscriptaddress,twocolumn,longbibliography]{revtex4-1}
\usepackage[T1]{fontenc}
\usepackage{amsmath}
\usepackage{amssymb}
\usepackage{mathtools}
\usepackage[normalem]{ulem}
\usepackage{graphicx}
\usepackage{siunitx}
\usepackage{datetime}
\usepackage{color}
\usepackage{newtxtext,newtxmath}
\usepackage{microtype}
\usepackage{braket}
\usepackage{xr}
\usepackage{natbib}
\usepackage{hyperref}
\hypersetup{%
    bookmarksopen=false,
    bookmarksnumbered=true,
    pdfnewwindow=true,
    unicode=false,
    colorlinks=true,%
    citecolor=blue,
    linkcolor=black,
    urlcolor=blue,
    filecolor=blue
    }
\newcommand{\red}[1]{{#1}}
\newcommand{\vb}[1]{\boldsymbol{#1}}
\newcommand{\NN}{\mathcal{N}}
\newcommand{\mrm}[1]{\mathrm{#1}}
\DeclareMathOperator{\var}{var}
\DeclareMathOperator{\Erf}{Erf}
\begin{document}
\newcommand{\figdir}{.}
\newcommand{\freiburg}{Physikalisches Institut, Albert-Ludwigs-Universit\"{a}t Freiburg, Hermann-Herder-Stra{\ss}e 3, D-79104, Freiburg, Germany}
\newcommand{\usal}{Departamento de F\'isica Fundamental, Universidad de Salamanca, E-37008 Salamanca, Spain}
\newcommand{\eucor}{EUCOR Centre for Quantum Science and Quantum Computing, Albert-Ludwigs-Universit\"{a}t Freiburg, Hermann-Herder-Stra{\ss}e 3, D-79104, Freiburg, Germany}
\title{Chaos and ergodicity across the energy spectrum of interacting bosons}
\author{Lukas Pausch}
\affiliation{\freiburg}
\author{Edoardo G. Carnio}
\affiliation{\freiburg}
\affiliation{\eucor}
\author{Alberto Rodr\'iguez}
\email[]{argon@usal.es}
\affiliation{\usal}
\author{Andreas Buchleitner}
\email[]{a.buchleitner@physik.uni-freiburg.de}
\affiliation{\freiburg}
\affiliation{\eucor}

\begin{abstract}
	We identify the chaotic phase of the Bose-Hubbard Hamiltonian by the energy-resolved correlation between spectral features and structural changes of the associated eigenstates as exposed by their generalized fractal dimensions.
	The eigenvectors are shown to become ergodic in the thermodynamic limit, in the configuration space Fock basis, in which random matrix theory offers a remarkable description of their typical structure.
	The distributions of the generalized fractal dimensions, however, are ever more distinguishable from 
	random matrix theory as the Hilbert space dimension grows.
\end{abstract}
\maketitle
Ergodicity, understood as the ability of a system to dynamically explore, irrespective of its initial state, all possible configurations at given energy, 
is, in general, an exceedingly difficult to prove and rather rare property, at the classical and quantum level \cite{lichtenberg83,ozorio88,ott93}.
On the quantum side, safe ground is established by (intrinsically ergodic \cite{Pandey1979}) random matrix theory (RMT), 
which describes systems with classically strictly chaotic (``K-systems'' \cite{lichtenberg83,ozorio88,bohigas89,ott93}) 
dynamics \cite{bgs84}. 
RMT predictions for energy spectra and eigenstates \cite{Berry1977,delande1986} define popular benchmarks to certify ergodicity \cite{Khaymovich2019,DeTomasi2020}. 

Ergodicity can, however, emerge on widely variable time scales, hinging on finer structures of phase space, and,
at the quantum level, on the effective coarse graining thereof induced by the finite size of $\hslash$ \cite{Izrailev1990}. 
Since the majority of dynamical systems features mixed 
rather than strictly chaotic dynamics \cite{geisel86,Bohigas1993,abu1995,ketzmerick2000,Hiller2006,Michailidis2020}, 
one therefore expects detectable deviations from RMT ergodicity \cite{Weidenmuller2009,Backer2019}, in particular at the level of the eigenvectors' structural properties---
which reflect the underlying phase space structure \cite{geisel86,Bohigas1993,abu1995,ketzmerick2000,Carvalho2004,Hiller2006}. 
This holds on the level of single as well as of many-body quantum systems,
with engineered ensembles of ultracold atoms \cite{Gericke2008,Karski2009,Gemelke2009,Sherson2010,Bakr2010,Meinert2016a} as a modern playground:
Notably, interacting bosons on a regular lattice provide a
paradigmatic experimental setting to explore the questions above \cite{Ronzheimer2013a,meinert2014,Preiss2015,Islam2015,Kaufman2016}; 
they feature chaos on the level of spectral \cite{Kolovsky2004,Kollath2010,Dubertrand2016,Fischer2016a} 
and eigenvector properties \cite{Kolovsky2004,Kollath2010,Beugeling2014,Beugeling2015,Beugeling2015c,Beugeling2018}
as well as quench dynamics \cite{abu2003,Sorg2014,Kollath2007,Roux2010,Biroli2010b}.

\begin{figure*}
	\centering
	\includegraphics[width=\textwidth]{\figdir/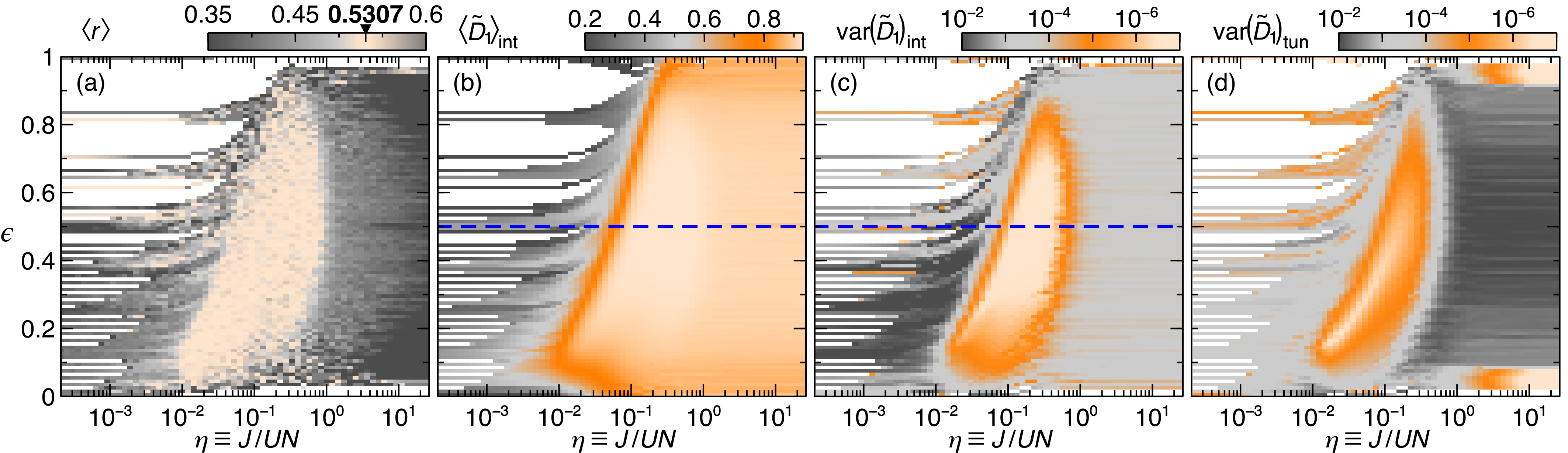}
	\caption{Evolution of $\langle r\rangle$ \red{(a)}, $\langle \tilde{D}_1\rangle$ \red{in the eigenbasis of $H_\mathrm{int}$ (b)} and $\var(\tilde{D}_1)$ \red{in the eigenbases of $H_\mathrm{int}$ (c) and $H_\mathrm{tun}$ (d)}, as functions of $\eta$ and energy $\epsilon = (E-E_\mrm{min})/(E_\mrm{max}-E_\mrm{min})$, for the irreducible Hilbert subspace of size $\mathcal{N}=55\,898$ 
		\red{with} $Q=0$ and $\pi=-1$, for $N=L=12$ with PBC.
		The spectrum was obtained for $75$ equally spaced values of $\log_{10}(J/U)\in[-2.92,3]$, and divided into 100 bins of equal width along the $\epsilon$ axis. 
		The value $\langle r\rangle_\mrm{GOE}$  is highlighted over the left color bar.
		Blue dashed lines mark the value $\epsilon = 0.5$ considered in Fig.~\ref{fig:D1vsJUN}.}
	\label{fig:en_dens_INT}
\end{figure*}

Here we consider the one-dimensional Bose-Hubbard Hamiltonian (BHH) and combine state-of-the-art numerical
simulations 
with analytical calculations to
establish a so far missing integral picture of its chaotic and non-ergodic  phases, providing deeper insight into the concept of chaos and ergodicity in the quantum realm.
We demonstrate
that (i) the energy-resolved chaotic phase is signalled by a clear correlation between spectral features and eigenstate structural changes captured by generalized fractal dimensions (GFD) (cf.~Fig.~\ref{fig:en_dens_INT}), whose fluctuations 
seem to be qualitatively basis-independent,
(ii) a non-ergodic phase persists in the thermodynamic limit, \red{as a function of a scaled tunneling strength,}
(iii) eigenvectors within the chaotic phase become ergodic in the thermodynamic limit in the configuration space Fock basis, 
where RMT provides a remarkable description of the eigenstates' typical (i.e., most probable) GFD, 
(iv) despite such agreement,
according to the GFD distributions 
BHH and RMT depart from each other in an unequivocal statistical sense 
with increasing size of Hilbert space.
This implies that the fluctuations of the eigenstates' structure along the path to ergodicity (even if it be arbitrarily close to RMT at a coarse-grained level) contain statistically robust fingerprints of the specific underlying Hamiltonian.

In terms of standard bosonic operators associated with $L$ Wannier spatial modes, 
the BHH \cite{Lewenstein2007,Cazalilla2011,Krutitsky2016} is the sum of
a tunneling and a local interaction Hamiltonian with respective strengths $J$ and $U$, 
\begin{eqnarray}
H_\mathrm{tun} &=& -J\sum_{k}(b^\dagger_k b_{k+1} + b^\dagger_{k+1} b_k), \label{eq:Htun}\\
H_\mathrm{int} &=& \frac{U}{2} \sum_k n_k \left(n_k -1\right). \label{eq:Hint}
\end{eqnarray}
The BHH exhibits a $Z_2$ symmetry 
under the reflection operation ($\Pi$) about the center of the lattice.
In the presence of periodic boundary conditions (PBC), the BHH additionally has translational symmetry, and  
Hilbert space can be decomposed into $L$ irreducible blocks distinguished by 
the center-of-mass quasimomentum $Q$. The $Q=0$ block further disjoins
into symmetric ($\pi=+1$) and antisymmetric ($\pi=-1$) subspaces. 
For hard-wall boundary conditions (HWBC), the latter $\pi$-division applies to the full Hilbert space.

Both $H_\mathrm{tun}$ and $H_\mathrm{int}$ are integrable and analytically solvable in appropriate Fock bases.
The eigenvectors of the interaction term are the Fock states of the on-site number operators, $\ket{\boldsymbol{n}}\equiv\ket{n_1,\ldots,n_L}$, with $||\vb{n}||_1=N$,
where $N$ is the number of bosons. 
The eigenvectors of $H_\mathrm{tun}$ follow from the Fock states of
number operators of spatially delocalized plane-wave or standing-wave modes, for PBC or HWBC, respectively. 

The competition between tunneling and interaction makes the BHH  non-integrable:
For comparable $J$ and $U$, it exhibits spectral chaos \cite{Kolovsky2004,Kollath2010,Dubertrand2016,Fischer2016a}, identified by 
short-range spectral measures in accord with the Gaussian orthogonal ensemble (GOE) of RMT. 
This may be traced back to the underlying classical Hamiltonian
\cite{Hiller2006,Hiller2009,Dubertrand2016}, whose dynamics are governed by 
the scaled energy $H/UN^2$ and the scaled tunneling strength $\eta\equiv J/UN$. In the quantum system, 
one therefore expects $\eta$ to control the emergence of chaos in sufficiently dense spectral regions. 

We numerically analyze the BHH at unit filling ($N=L$): 
Eigenstates around chosen energy targets \cite{Pietracaprina2018,petsc-user-ref,slepc} as well as full spectra, 
scaled as $\epsilon \equiv (E-E_\mrm{min})/(E_\mrm{max}-E_\mrm{min})\in[0,1]$, enabling the juxtaposition of results for different $N$ and $\eta$,
are obtained by exact diagonalization.
Since the form of $H_\mathrm{tun}$ and $H_\mathrm{int}$ reveals that $E_\mrm{max}-E_\mrm{min}\sim UN^2$ for large $N$, 
$\epsilon$ effectively provides the classically scaled energy.
Short-range statistical features of the spectrum are best captured by the level spacing ratios \cite{Oganesyan2007,Atas2013c}, $r_n = \min(s_{n+1}/s_n, s_n/s_{n+1})$, where $s_n = E_{n+1}-E_n$ is the $n$-th level spacing. 
The distributions of $r$ are known approximately analytically for Gaussian random matrix ensembles and accessible numerically without 
further unfolding procedures, e.g., $\langle r \rangle_\mathrm{GOE}\approx 0.5307$ \cite{Atas2013c}.

The eigenstate structure of generic many-body Hamiltonians in Hilbert space
exhibits multifractal complexity \cite{Stephan2009,Stephan2010,Stephan2011,Atas2012,Atas2014,Luitz2014,Luitz2015,Misguich2017,Lindinger2019,Mace2019,Luitz2020}, and is conveniently described by finite-size generalized fractal dimensions (GFD) \cite{Rodriguez2011,Lindinger2019},
\begin{equation}
\tilde{D}_q\equiv \frac{1}{1-q} \log_\mathcal{N} R_q,\quad \mathrm{with}\;\; R_q = \sum_{\alpha} \left| \psi_{\alpha}\right|^{2 q},\;  q\in\mathbb{R}^+,
\label{eq:Dq}
\end{equation}
for eigenvectors with amplitudes $\psi_{\alpha}$ in a given orthonormal basis of size $\mathcal{N}$.
The eigenvector moments are expected to scale asymptotically as $R_q\sim \mathcal{N}^{-(q-1)D_q}$, where the dimensions $D_q\equiv \lim_{\mathcal{N}\to\infty}\tilde{D}_q$ decide whether the state is localized ($D_q=0$ for $q\geqslant 1$ \cite{note1}), multifractal (extended non-ergodic; $q$-dependent $0<D_q<1$), or ergodic ($D_q=1$ for all $q$), 
in the chosen expansion basis.
Consequently, the support of ergodic eigenstates---e.g., the eigenvectors of the Wigner-Dyson 
RMT ensembles \cite{Mirlin1999}---scales asymptotically as the full Hilbert space.
Among all GFD, we focus on $\tilde{D}_1$, 
governing the scaling of the Shannon 
entropy of  $\{|\psi_\alpha|^2\}$, $\tilde{D}_2$, determining the scaling of the eigenstate's inverse participation ratio, and
$\tilde{D}_\infty = -\log_\mathcal{N} \max_\alpha |\psi_\alpha|^2$, 
unveiling the extreme statistics of the state's intensities.

We first analyze the connection between spectral chaos and the eigenstates' GFD.
In Fig.~\ref{fig:en_dens_INT}\red{(a)-(c)}, we show the evolution of $\langle r \rangle$, $\langle \tilde{D}_1\rangle$, and $\var(\tilde{D}_1)$, as functions of scaled energy $\epsilon$ and scaled tunneling strength $\eta$, for $N=12$ and PBC (subspace $Q=0$, $\pi=-1$), 
evaluated in the eigenbasis of $H_\mathrm{int}$. The $\epsilon$ spectrum is divided into 100 bins of equal width; 
mean values and variances are computed from eigenvalues and eigenvectors falling into each bin.
Energy-resolved density plots expose the coarse-grained level dynamics of the system: 
Heavily degenerate manifolds of $H_\mathrm{int}$ fan out as $\eta$ increases, 
overlap, and then form a bulk region massively populated by avoided crossings (observable upon finer inspection \cite{pausch}), 
which eventually dissolves as the levels reorganize into the bands allowed by $H_\mathrm{tun}$, for $\eta\gg 1$.
We identify a slightly bent oval region of spectral chaos, centered around $\eta \simeq 0.1$ and extending over $0.1\lesssim \epsilon \lesssim 0.9$,  where $\langle r \rangle$ attains the GOE value.
This region remains visible after averaging $r$ over a large portion of the bulk spectrum, 
even without resolving the $\Pi$-symmetry \cite{SM}. 
The onset of spectral chaos correlates with a sudden increase in the eigenvectors' GFD, which reach maximum values within the spectral chaos region, as demonstrated for $\langle \tilde{D}_1\rangle$. 
Strictly simultaneously, the energy-resolved GFD variance undergoes a dramatic reduction by several orders of magnitude.
This behavior is also revealed by $\tilde{D}_2$ and $\tilde{D}_\infty$,
and qualitatively the same in any irreducible subspace, also for HWBC.
The chaotic regime can therefore be identified by the unambiguous correlation between spectral features and 
structural changes of eigenstates, which, as revealed by the GFD, tend to homogenize their delocalization in Hilbert space, 
irrespective of their energy.

\begin{figure}
	\centering
	\includegraphics[width=\columnwidth]{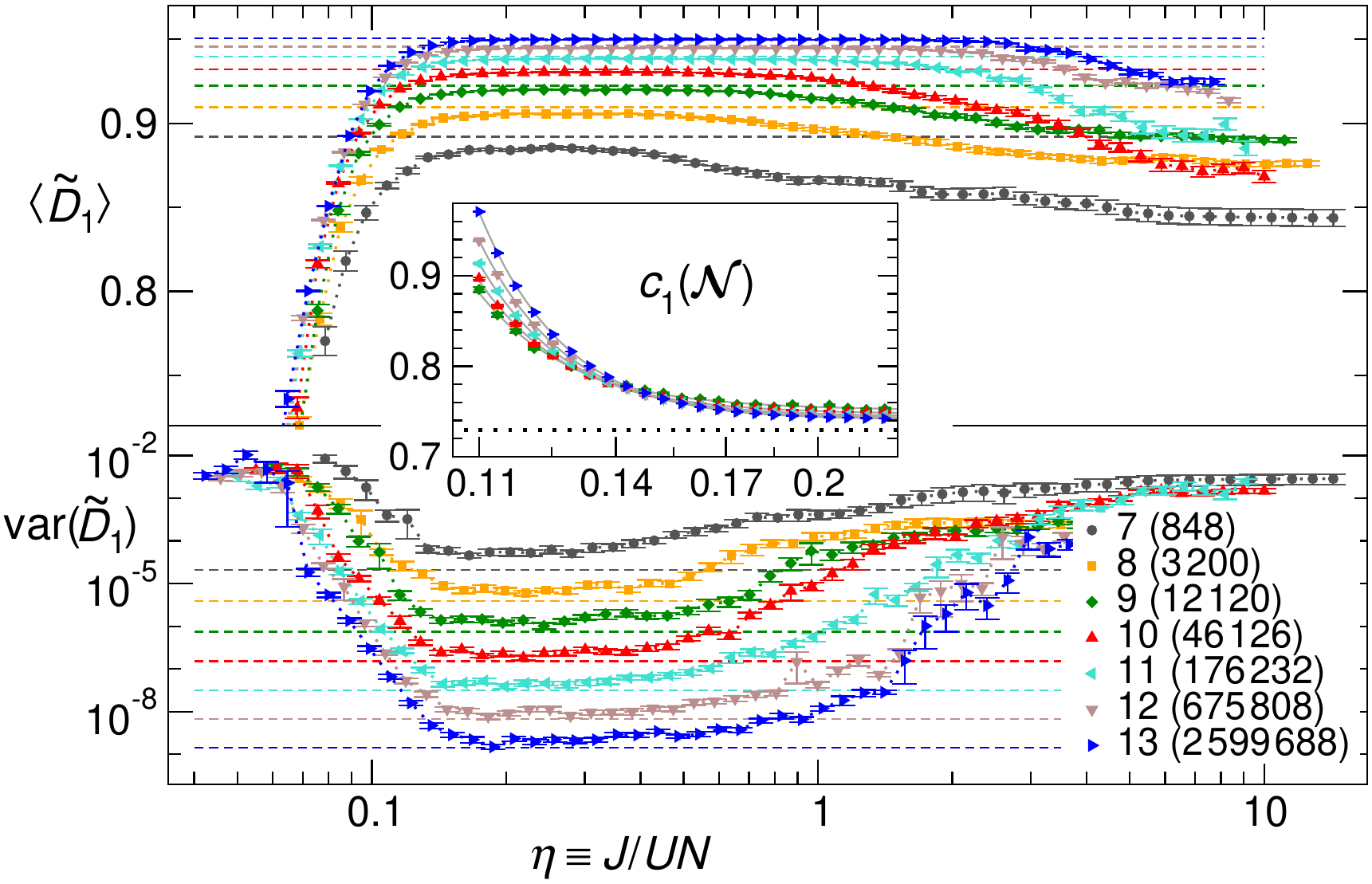}
	\caption{Evolution of $\langle \tilde{D}_1\rangle$ (top) and $\var(\tilde{D}_1)$ (bottom) at $\epsilon=0.5$ versus $\eta$, for varying values of $L$ and size $(\mathcal{N})$ (as indicated by the legend) of the  subspace 
		spanned by the $\pi=-1$ eigenstates of $H_\textrm{int}$ with HWBC.
		Each data point results from the analysis of the $100$ BHH eigenvectors closest to $\epsilon=0.5$.
		Corresponding GOE values
		are indicated by dashed lines. The inset shows the behavior of $c_1(\mathcal{N})=(1-\langle \tilde{D}_1\rangle)\ln\NN$ versus $\eta$ around the crossover region (solid lines are guides to the eye). The horizontal dotted line marks the GOE value of $c_1(\mathcal{N}\to\infty)$.}
	\label{fig:D1vsJUN}
\end{figure}

To elucidate the eigenstates' structural dependence on Hilbert space's size, 
Fig.~\ref{fig:D1vsJUN} shows mean and variance of $\tilde{D}_1$, for fixed $\epsilon=0.5$ 
(where the density of states is maximum once spectral chaos kicks in), versus $\eta$, for increasing size (up to $\mathcal{N}\approx 2.6\times 10^6$) of the $\pi=-1$ subspace with HWBC. 
$\langle\tilde{D}_1\rangle$ registers a surge around $\eta=0.1$, and reaches a maximum that develops into a distinct plateau, extending towards larger $\eta$ for increasing $L$. (Also $\langle r\rangle$ exhibits plateau broadening 
at $\epsilon=0.5$ \cite{pausch}.) This behavior is mirrored by the drastic (ever bigger, with increasing $L$) drop of $\var(\tilde{D}_1)$, 
with plateaux at its minima. 
Note that the plateau values of $\langle\tilde{D}_1\rangle$ and $\var(\tilde{D}_1)$ 
agree well with those expected for GOE eigenvectors, 
indicated by dashed lines in Fig.~\ref{fig:D1vsJUN}.
The same is qualitatively observed for $q=2,\infty$, 
other irreducible subspaces, and PBC.
The onset of the plateaux 
appears system size independent in terms of $\eta$ \cite{SM}, 
confirming the relevance of the classically scaled tunneling strength.

\begin{figure}
	\centering
	\includegraphics[width=\columnwidth]{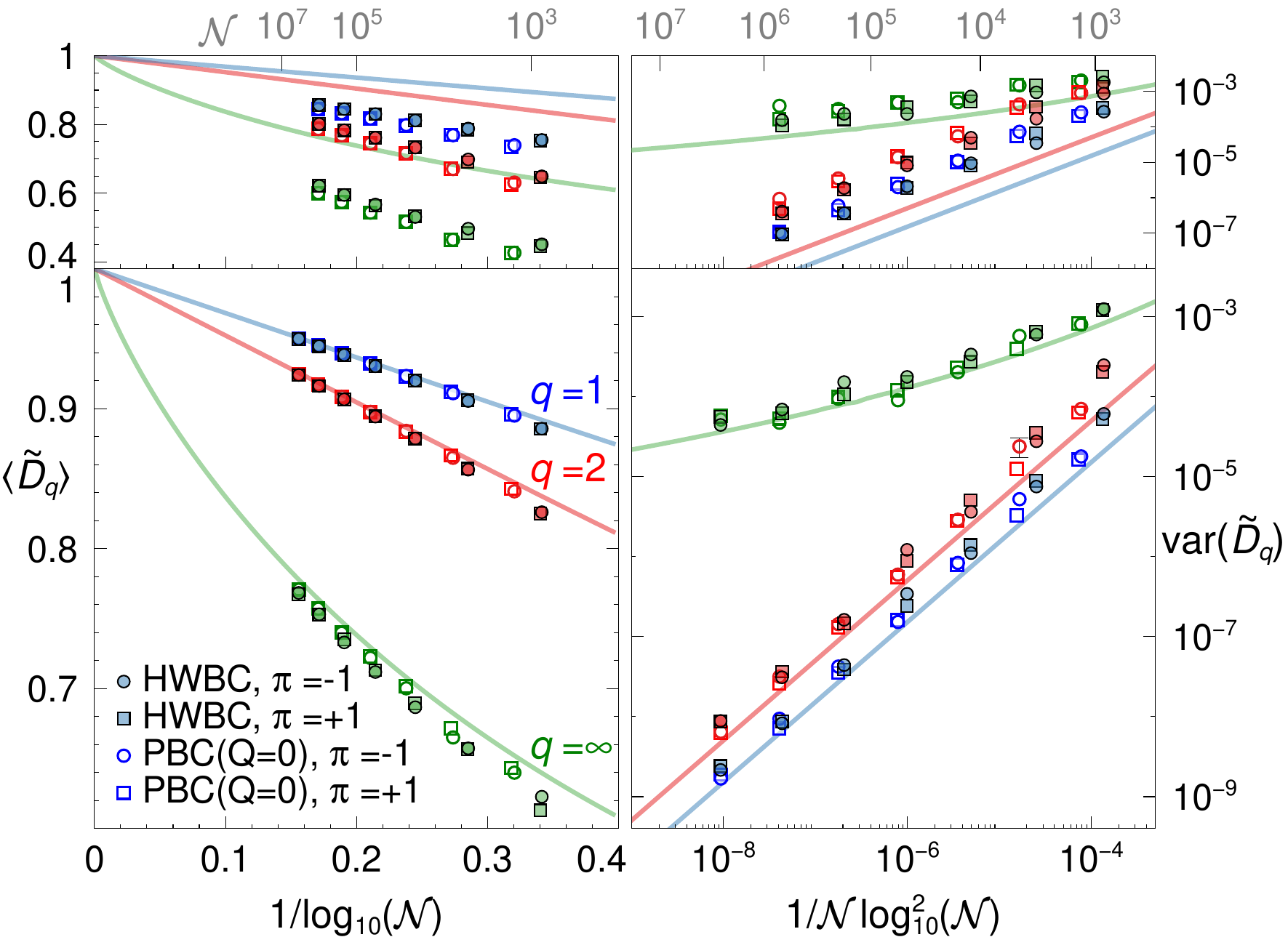}
	\caption{Average and variance of $\tilde{D}_1$, $\tilde{D}_2$, and $\tilde{D}_\infty$, at $\eta=0.25$ and $\epsilon=0.5$, versus size $\mathcal{N}$ of four Hilbert subspaces 
		(distinguished by symbols as indicated; each data point involves 100 eigenstates as in Fig. \ref{fig:D1vsJUN}). Lower (upper) panels correspond to the analysis in the eigenbasis of $H_\textrm{int}$ ($H_\textrm{tun}$).
		Solid lines show GOE predictions. Whenever not shown, errors are contained within symbol size.}
	\label{fig:ScalingDq}
\end{figure}

To shed further light on the GFD asymptotics within the chaotic region, 
the lower panels of Fig.~\ref{fig:ScalingDq} show $\langle\tilde{D}_q\rangle$ and $\var(\tilde{D}_q)$ at $\epsilon=0.5$ and $\eta=0.25$, for increasing 
$\mathcal{N}$ of four irreducible subspaces, 
evaluated in the corresponding eigenbases of $H_\mathrm{int}$. 
The results are compared against the GOE values,  
which, using known distributions \cite{Haake2004} and 
extreme statistics  \cite{Lakshminarayan2008}, can be estimated analytically \cite{SM}. 
We find, asymptotically,
\begin{eqnarray}
\langle \tilde{D}_1\rangle_\mathrm{GOE} &=& 1-\frac{1}{\ln\mathcal{N}}\left[2-\gamma-\ln 2 - \frac{1}{\mathcal{N}} 
+O\left(\mathcal{N}^{-2}\right)\right], \\
\langle \tilde{D}_\infty\rangle_\mathrm{GOE} &=&1-\frac{\ln (2\ln\mathcal{N})}{\ln\mathcal{N}} +O\left(\ln\ln\mathcal{N}/\ln^2\mathcal{N}\right),
\end{eqnarray}
where $\gamma$ is Euler's constant, and
\begin{eqnarray}
\var(\tilde{D}_1)_\mathrm{GOE}&=&\frac{1}{\ln^2\mathcal{N}}\left[\frac{3 \pi ^2-28}{2 \mathcal{N}}
+O\left(\mathcal{N}^{-2}\right)\right], \\
\var(\tilde{D}_\infty)_\mathrm{GOE}&\sim& \ln^{-4}\mathcal{N}.
\end{eqnarray}
For $q=2$, we compare the results 
to the ensemble-averaged GFD, $\langle\tilde{D}_q^{(\textrm{ens})}\rangle_\mathrm{GOE}=\log_{\NN}\langle R_q\rangle/(1-q)$, instead \cite{Backer2019}\cite{SM}, 
with finite-size corrections found identical (up to coefficients) 
with those for $\tilde{D}_1$. 
As shown in Fig.~\ref{fig:ScalingDq}, the GFD, as well as $\var(\tilde{D}_q)$, 
in the eigenbasis of $H_\mathrm{int}$ 
quickly approach GOE values, 
independently of subspace or boundary conditions 
(for the largest $\mathcal{N}$ shown, $\langle\tilde{D}_1\rangle_\mathrm{GOE}-\langle\tilde{D}_1\rangle =8\times10^{-4}$).
The dominant finite-size correction seems to have the same $\NN$ dependence for BHH as for GOE eigenvectors, and
the GFD show clear evidence of converging to $1$ in the thermodynamic limit (as the corresponding variance vanishes). We therefore conclude that the BHH eigenvectors in the chaotic regime become ergodic in the eigenbasis of $H_\mathrm{int}$ in the thermodynamic limit.

Hence, as $\mathcal{N}\to\infty$, the plateau value of $\langle\tilde{D}_1\rangle$ in Fig.~\ref{fig:D1vsJUN} approaches $1$, and, although 
the crossover into the chaotic region 
becomes more pronounced with larger $L$, we cannot definitely determine whether it turns into a sharp transition (i.e., a discontinuity of the derivative with respect to $\eta$) 
or remains smooth and differentiable. 
The transition features a standard scaling behavior \cite{Rodriguez2011,Luitz2014,Luitz2015,Mace2019}
in terms of $c_1(\mathcal{N})\equiv (1-\langle\tilde{D}_1\rangle)\ln\mathcal{N}$: For increasing $L$, 
$c_1$ is unbounded in the non-ergodic phase (where $\langle\tilde{D}_1\rangle<1$, i.e., the eigenstates are generically multifractal), 
and decreases to converge to a constant value in the chaotic phase if the dominant finite-size correction is $\ln^{-1}\mathcal{N}$. That is indeed the behavior observed numerically 
(inset of Fig.~\ref{fig:D1vsJUN}),
which confirms that the non-ergodic phase for small $\eta$ persists in the thermodynamic limit.
Given the lack of analytical information, we
refrain from detailed finite-size scaling analyses on $c_1$. 
Nonetheless, close inspection of the tendency 
of the data locates the transition/crossover, at $\epsilon=0.5$, in the thermodynamic limit within the region $\eta\in[0.15,0.2]$ 
to a reasonable level of confidence. 
The plateaux's right termination points 
show no hint of reaching a finite asymptotic value for increasing $L$, an absence 
less pronounced for PBC \cite{SM}. Although it is appealing to think that an 
infinitesimal interaction suffices to induce ergodicity in the thermodynamic limit (as discussed for fermions \cite{Neuenhahn2012}), and hence that the chaotic phase might have no upper $\eta$ 
limit (the point $\eta=\infty$ then being a discontinuity), further investigation is necessary to verify such hypothesis.

\begin{figure}
	\centering
	\includegraphics[width=\columnwidth]{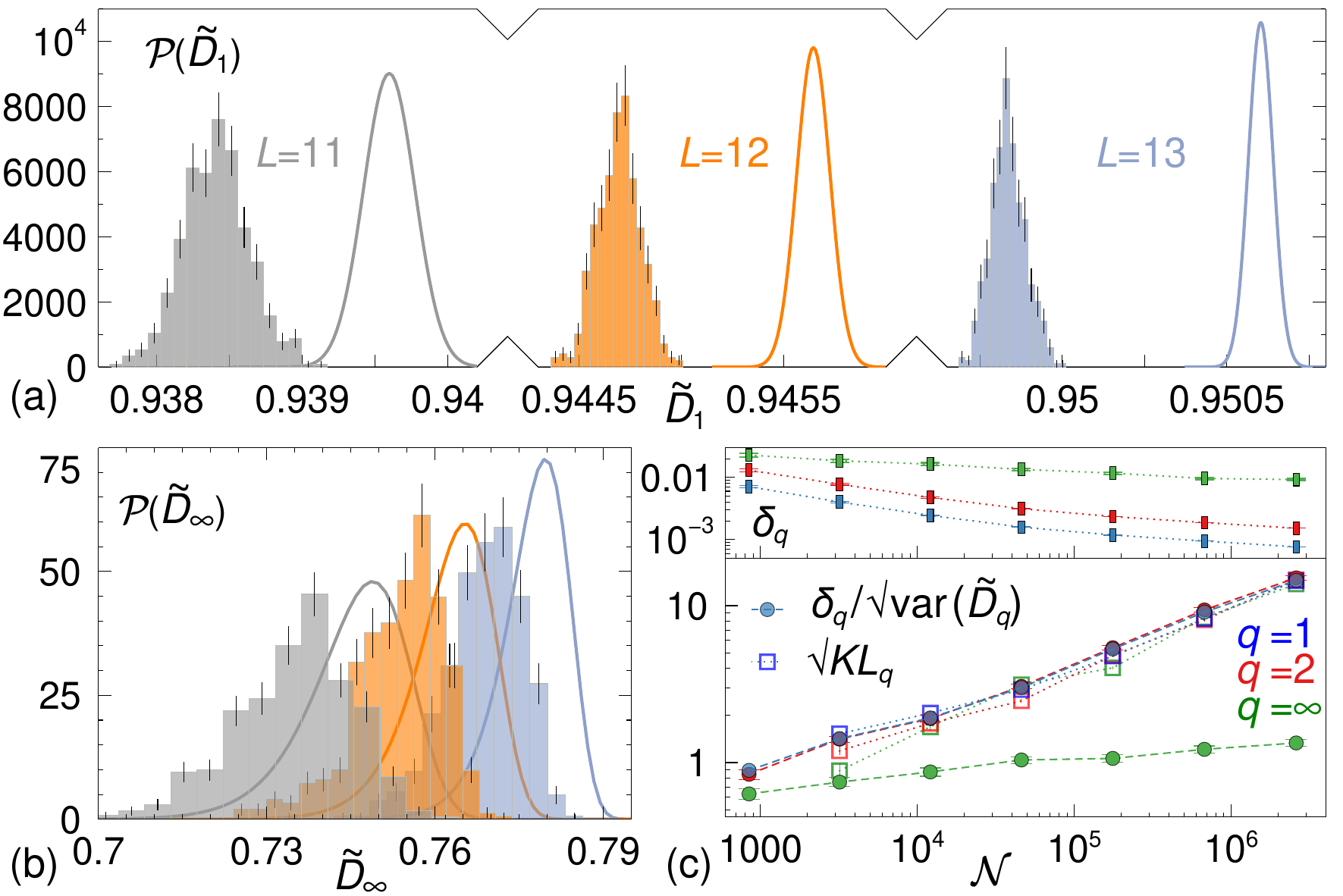}
	\caption{Evolution of the probability density function of $\tilde{D}_q$ with increasing size $\mathcal{N}$ of the subspace spanned by the $\pi=-1$ eigenstates of $H_\textrm{int}$ with HWBC. Panels (a) and (b) display the distributions of $\tilde{D}_1$ and $\tilde{D}_\infty$, respectively, for the indicated $L$ values. Each histogram comprises 500 eigenstates at $\epsilon=0.5$ in the chaotic domain (100 eigenstates $\times$ five values of $\eta\in[0.25,0.38]$).
		The $\tilde{D}_1$ distribution for $L=11$ ($L=12$) is normalized to 4 (2) for better visualization. Solid lines show GOE distributions \cite{SM},
		the distance to which 
		is evaluated in panels (c), via the difference $\delta_q$ of the means (upper plot), the renormalized difference $\delta_q/\sqrt{\var(\tilde D_q)}$ and the Kullback-Leibler divergence (lower plot).}
	\label{fig:PDFs}
\end{figure}

Given the demonstrated quality of RMT predictions, one may naively conclude that, at the level of simple eigenvector observables such as 
Hilbert space (de)localization captured by GFD, as $L$ grows the BHH unequivocally assumes 
universal RMT behavior within its chaotic phase. But a detailed inspection indicates otherwise: 
Analysis of the full GFD distributions 
in Fig.~\ref{fig:PDFs} reveals that, although the first and second moments 
approach the GOE values, the distributions become 
{\em more distinguishable} from GOE 
as $L$ increases. The distance between BHH and GOE distributions is quantified in Fig.~\ref{fig:PDFs}(c) using
the square root of the Kullback-Leibler divergence (relative entropy) \cite{KullbackLeibler,InformationTheory}, $\sqrt{KL_q}$, and
$d_q(\mathcal{N})\equiv\delta_q/\sqrt{\var(\tilde D_q)}$, 
where $\delta_q\equiv \langle\tilde{D}_q\rangle_\mathrm{GOE}-\langle\tilde{D}_q\rangle$.
Both of these measures increase with $L$ for $q=1,2,\infty$, demonstrating that, even at the level of the GFD, the two models depart from each other in a statistically unambiguous way: For $\mathcal{N}\gtrsim10^6$ ($L\geqslant 13,15$, depending on boundary conditions) the typical 
$\tilde{D}_{1,2}$ 
lies more than $10\sigma$ away from the most probable GOE value.
The distance between the GFD distributions provides valuable information on the finite-size corrections for BHH eigenvectors. 
Since $\var(\tilde D_{1,2})\sim 1/\mathcal{N}\ln^2\mathcal{N}$, 
the growth of $d_{1,2}(\mathcal{N})$ with $\NN$ entails that $\left\langle\tilde{D}_{1,2}\right\rangle$ and $\left\langle\tilde{D}_{1,2}\right\rangle_\mrm{GOE}$ differ at terms 
decaying slower than $1/\sqrt{\NN}\ln \NN$. 
A data inspection indicates $d_{1,2}(\mathcal{N})\sim \sqrt{\mathcal{N}}/\ln\mathcal{N}$ as the most likely behavior for large $\mathcal{N}$, implying that $\left\langle\tilde{D}_{1,2}\right\rangle$ bear a $1/\ln^2 \NN$ subleading correction \cite{SM}.
Note that, for non-overlapping Gaussian distributions of similar width, $\sqrt{KL_q}$ is equivalent to $d_q(\mathcal{N})$. 
Hence, comparison of these two quantities also provides the distributions' deviation from Gaussianity, 
as manifestly visible for $q=\infty$.

We finally address the chaotic eigenstates' features' dependence 
on the expansion basis. Although the GFD are naturally basis dependent, the eigenstates' ergodic character 
in the thermodynamic limit suggests some degree of invariance under rotations.
An analysis performed in the eigenbasis of $H_\mathrm{tun}$, instead of $H_\mathrm{int}$, reveals the same qualitative behavior of the energy-resolved $\var(\tilde{D}_q)$
\red{as demonstrated in Fig.~\ref{fig:en_dens_INT}(d): The GFD fluctuations are strongly suppressed, by several orders of magnitude, and undergo an ever bigger drop for larger system size, within the same parameter region as observed in the $H_\mathrm{int}$ basis and in the averaged $r$ statistic.}
Nonetheless, in the eigenbasis of $H_\mathrm{tun}$, 
there is no clear identification of a $\langle \tilde{D}_q\rangle$ plateau in the chaotic region \cite{SM}, and the typical GFD are distant from the GOE values, 
\red{as shown in} 
the upper panels of Fig.~\ref{fig:ScalingDq}.
If the GFD in this
basis converge to the ergodic limit, too, this is a much slower process governed by stronger finite-size corrections.
Such basis dependence reflects the different dynamics that excited eigenstates of $H_\mathrm{int}$ or of $H_\mathrm{tun}$ will exhibit under the BHH unitary evolution: 
While the first 
display 
indications of chaos 
already in relatively small systems \cite{Sorg2014,Kaufman2016}, the second 
may be substantially dominated by finite-size/finite-time effects.

We provided an integral view on the chaotic and non-ergodic phases of the Bose-Hubbard Hamiltonian, 
established by an energy-resolved correlation between spectral features and eigenstate structural changes exposed by the typical values and fluctuations of 
generalized fractal dimensions. 
Our results suggest that 
GFD fluctuations are far more sensitive probes of emergent chaotic behavior than the GFD themselves, and  
may 
identify the chaotic phase in any non-trivial basis.
In the eigenbasis of the Hamiltonian's interaction part, 
the chaotic phase eigenvectors become 
ergodic in the thermodynamic limit, and 
are remarkably well described by RMT.
Yet, 
in terms of the GFD distributions, 
their path towards ergodicity 
turns increasingly more distinguishable from RMT for larger Hilbert spaces, which suggests a statistical handle to discriminate bona fide BHH dynamics 
in the limit of numerically intractable Hilbert space dimensions. This relates our present results to the field of the certification of distinctive rather than universal 
features of complex quantum systems \cite{Tichy2010,Aolita2015,Giordani2018,Zache2020}. 
Whether this distinct GFD statistics of BHH with respect to RMT 
can be traced down to unambiguously unique features of the underlying Hamiltonian, or, alternatively, 
accommodated by 
more sophisticated random matrix ensembles
\cite{Mon1975,Benet2003,Chavda2017}, awaits further scrutiny. 

\begin{acknowledgments}
	We thank J.D.~Urbina for helpful discussions, and acknowledge support by the state of Baden-W\"urttemberg through bwHPC and the German Research Foundation (DFG) through Grants No.~INST 40/467-1 FUGG (JUSTUS cluster), No.~INST 40/575-1 FUGG (JUSTUS 2 cluster),
	and No.~402552777.
	E.G.C.~acknowledges support from the Georg H.~Endress foundation.
	A.R.~acknowledges support by Universidad de Salamanca through a Grant USAL-C1 (No.~18K145).
\end{acknowledgments}
%
%
%
\newcommand{\TITLE}{Chaos and ergodicity across the energy spectrum of interacting bosons}
\title{Supplemental Material \\[2mm] \TITLE}
%
%
%
%
\newpage
\onecolumngrid
\begin{center}\large\bfseries Supplemental Material \\[2mm] \TITLE \end{center}
\vskip 4mm
\twocolumngrid
\renewcommand\theequation{S\arabic{equation}}
\renewcommand\thefigure{S\arabic{figure}}
\setcounter{figure}{0}
\setcounter{equation}{0}
\section{Further results on spectral features and eigenstate structure of the BHH}
Table \ref{tab:dimensions} lists the sizes of the different irreducible Hilbert spaces considered in our numerical analysis. It is worth noting that while in the basis of $H_\mathrm{int}$ the number of non-zero elements per row in the Hamiltonian matrix grows linearly with $L$, the sparsity is severely reduced in the basis of $H_\mathrm{tun}$, where the number of non-zeros scales as $L^3$, which makes the numerical treatment far more demanding. The analysis in the tunneling eigenbasis is restricted to $L\leqslant 12$ ($L\leqslant 14$) for hard-wall (periodic) boundary conditions.
\begin{table*}
	\begin{tabular}{c|ccccccccc}
		\hline\hline
		$L$ & 7 & 8 & 9 & 10 & 11 & 12 & 13 & 14 & 15 \\\hline
		HWBC, $\pi=-1$ & 848 & \SI{3200}{} & \SI{12120}{} & \SI{46126}{} & \SI{176232}{} & \SI{675808}{} & \SI{2599688}{} & & \\
		HWBC, $\pi=+1$ & 868 & \SI{3235}{} & \SI{12190}{} & \SI{46252}{} & \SI{176484}{} & \SI{676270}{} & \SI{2600612}{} & & \\
		PBC ($Q=0$), $\pi=-1$ &  &  &  \SI{1317}{} & \SI{4500}{} & \SI{15907}{} & \SI{55898}{} & \SI{199550}{} & \SI{714714}{} & \SI{2583586}{} \\
		PBC ($Q=0$), $\pi=+1$ &  &  &  \SI{1387}{} & \SI{4752}{} & \SI{16159}{} & \SI{56822}{} & \SI{200474}{} & \SI{718146}{} & \SI{2587018}{} \\\hline
	\end{tabular}
	\caption{Size of the analyzed irreducible Hilbert spaces as function of $L$.}
	\label{tab:dimensions}
\end{table*}

\begin{figure}
	\centering
	\includegraphics[width=\columnwidth]{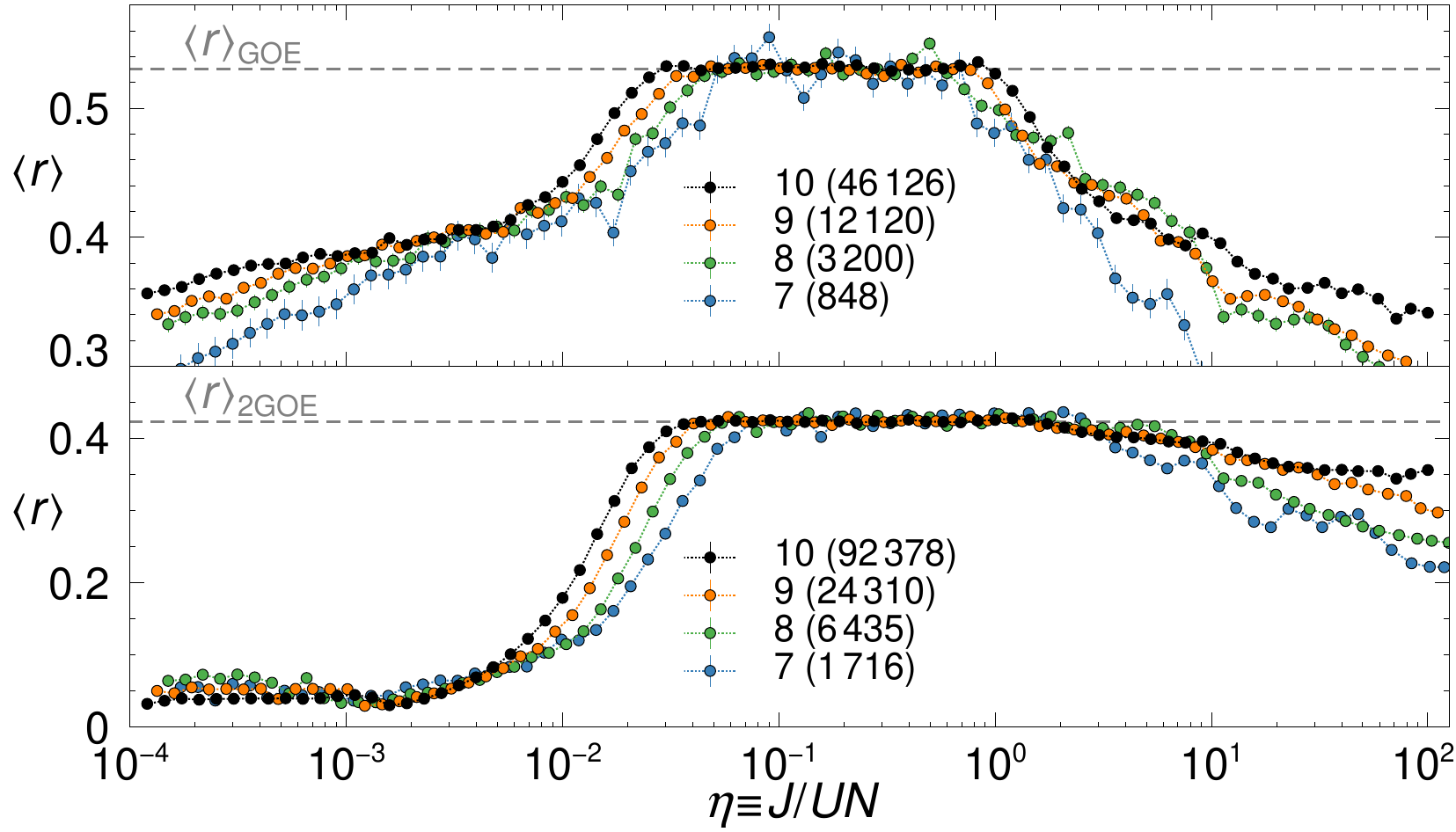}
	\caption{Average of $r$ over the inner 70\% of the eigenenergies 
		as a function of $\eta$ for varying values of $L$ and size $(\mathcal{N})$ (as indicated by the legend) of the reflection-antisymmetric subspace (top) as well as of the full Hilbert space 
		with hard-wall boundary conditions.
		Expected results for GOE eigen\red{values (top) and two superimposed GOE spectra (bottom)} are indicated by dashed lines. \red{$\left<r\right>_\mrm{2GOE}=0.4232\pm0.0002$ was obtained numerically by simulating 1000 realizations of the superimposed spectra of two $1000\times1000$ GOE random matrices and agrees well with recent analytical predictions \cite{Giraud2020}.} }
	\label{fig:aveRn}
\end{figure}

The average over the energy axis of the level statistics, as usually considered in the literature \cite{Kolovsky2004,Kollath2010,Dubertrand2016,Fischer2016a}, also reveals the existence of spectral chaos. 
Figure \ref{fig:aveRn} shows the level spacing ratio 
averaged over the inner 70\% of the eigenenergies 
as a function of $\eta\equiv J/UN$ for varying system sizes with hard-wall boundary conditions.
Agreement with the results expected for GOE eigenvalues is clearly observed within a range of the interaction strength which correlates with the behavior shown in Fig.~\ref{fig:en_dens_INT}. The region of spectral chaos is also visible even upon consideration of the full Hilbert space, i.e., without resolving the symmetric and antisymmetric subspaces induced by the reflection symmetry about the center of the chain.

\begin{figure}
	\centering
	\includegraphics[width=.95\columnwidth]{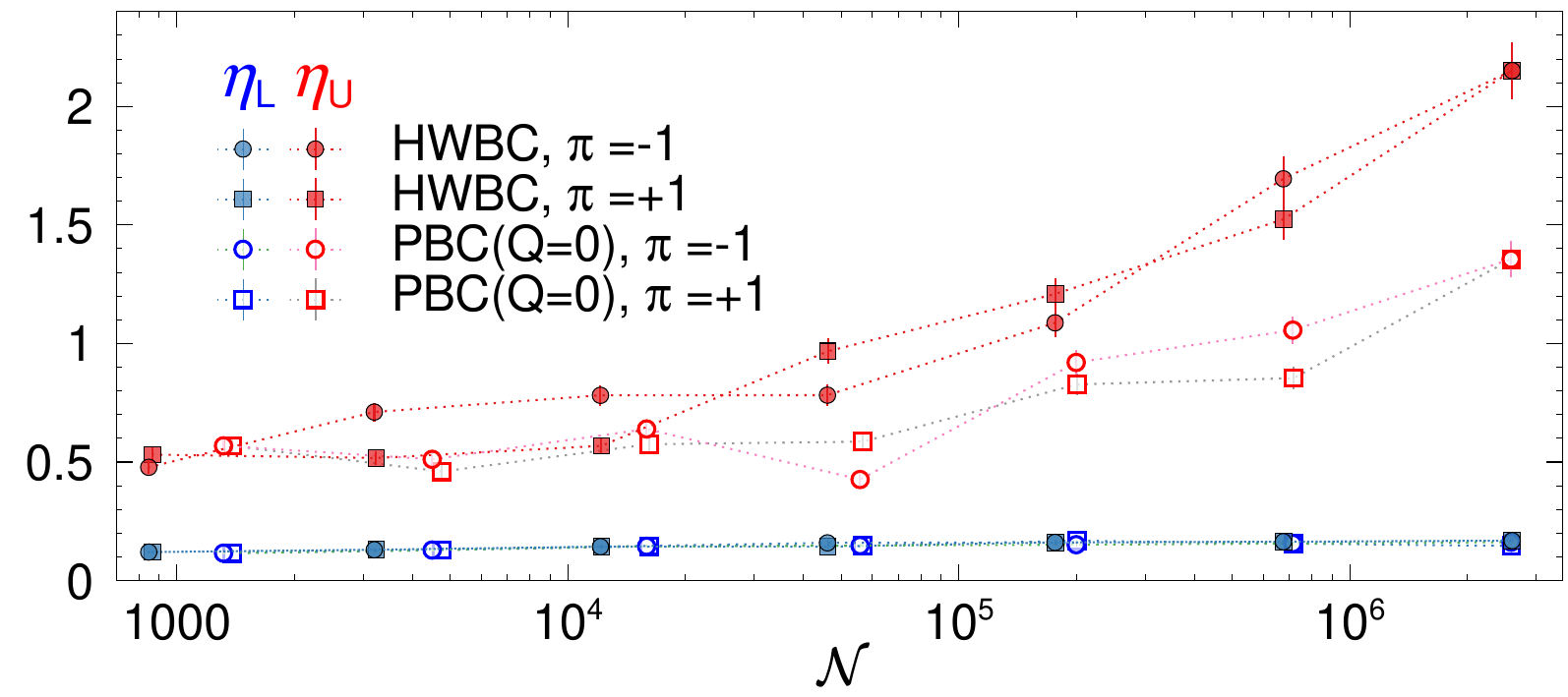}
	\caption{Estimation of the lower ($\eta_L$) and upper ($\eta_U$) limits of the $\langle \tilde{D}_1\rangle$ plateau at $\epsilon=0.5$ versus Hilbert space size, for four irreducible spaces resulting from the combination of translation and reflection symmetries, distinguished by different symbols as indicated in the plot.}
	\label{fig:ChaoticLimits}
\end{figure}

As can be seen in Fig.~\ref{fig:D1vsJUN} of the manuscript, the upper limit of the $\langle \tilde{D}_1\rangle$ plateau seems to keep increasing for larger $L$.  To quantify that behavior, we estimate the lower and upper limits of the plateaux as the positions where the difference between the typical GFD value and the corresponding GOE value is twice the minimum difference between these two,  
and show their evolution as functions of Hilbert space size for different boundary conditions in Fig.~\ref{fig:ChaoticLimits}. While the lower limit, $\eta_L$, converges to a value in the region $[0.15,0.2]$ for all the cases considered, the upper limit, $\eta_U$, does not exhibit an asymptotic saturation (especially for HWBC) in the accessible range of $\mathcal{N}$.

\begin{figure}
	\centering
	\includegraphics[width=\columnwidth]{\figdir/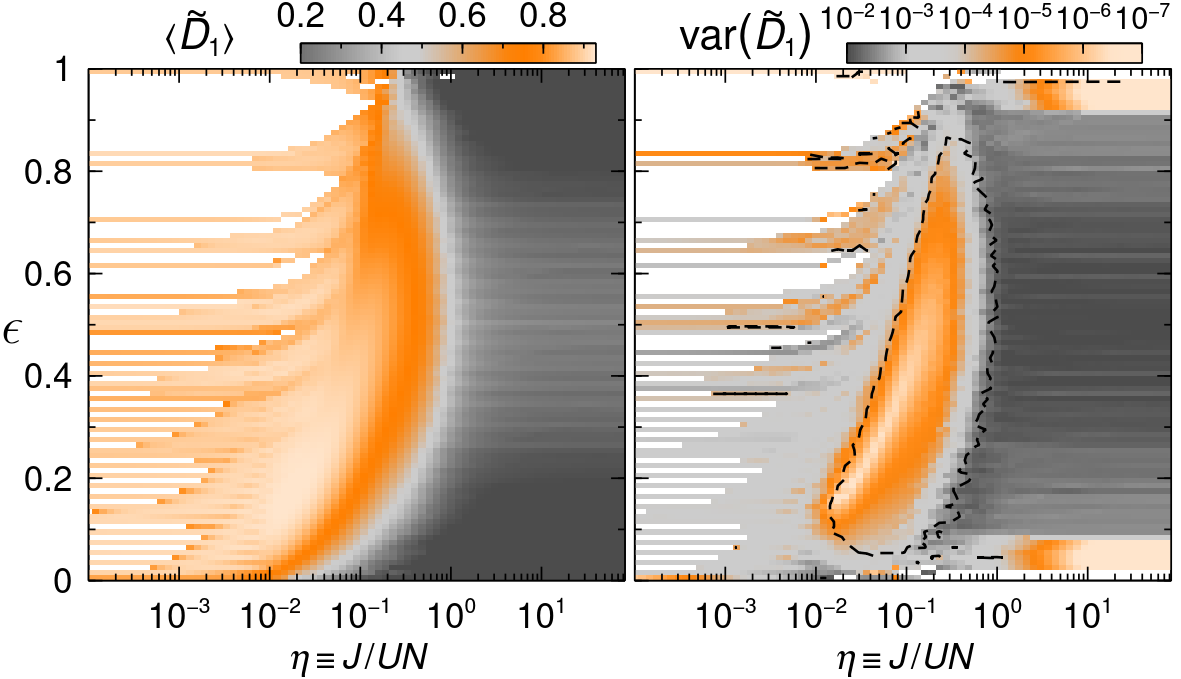}
	\caption{Evolution of $\langle \tilde{D}_1\rangle$ (left) and $\var(\tilde{D}_1)$ (right) as functions of $\eta$ and rescaled energy $\epsilon = (E-E_\mrm{min})/(E_\mrm{max}-E_\mrm{min})$, for the irreducible subspace of size $\mathcal{N}=55\,898$ spanned by the $Q=0$ and $\pi=-1$ eigenbasis of $H_\mrm{tun}$ for $N=L=12$ with PBC (cf.~Fig.~\ref{fig:en_dens_INT} in manuscript). The black dashed line in the right panel indicates the contour $\var(\tilde{D}_1)=\SI{5e-5}{}$ for the analysis in the eigenbasis of $H_\mrm{int}$, shown in Fig.~\ref{fig:en_dens_INT} of the manuscript. \red{The color scales for  $\langle\tilde{D}_1\rangle$ and $\var(\tilde{D}_1)$ are exactly the same as those used in Fig.~\ref{fig:en_dens_INT} of the manuscript.}} 
	\label{fig:en_dens_TUN}
\end{figure}

\begin{figure}
	\centering
	\includegraphics[width=\columnwidth]{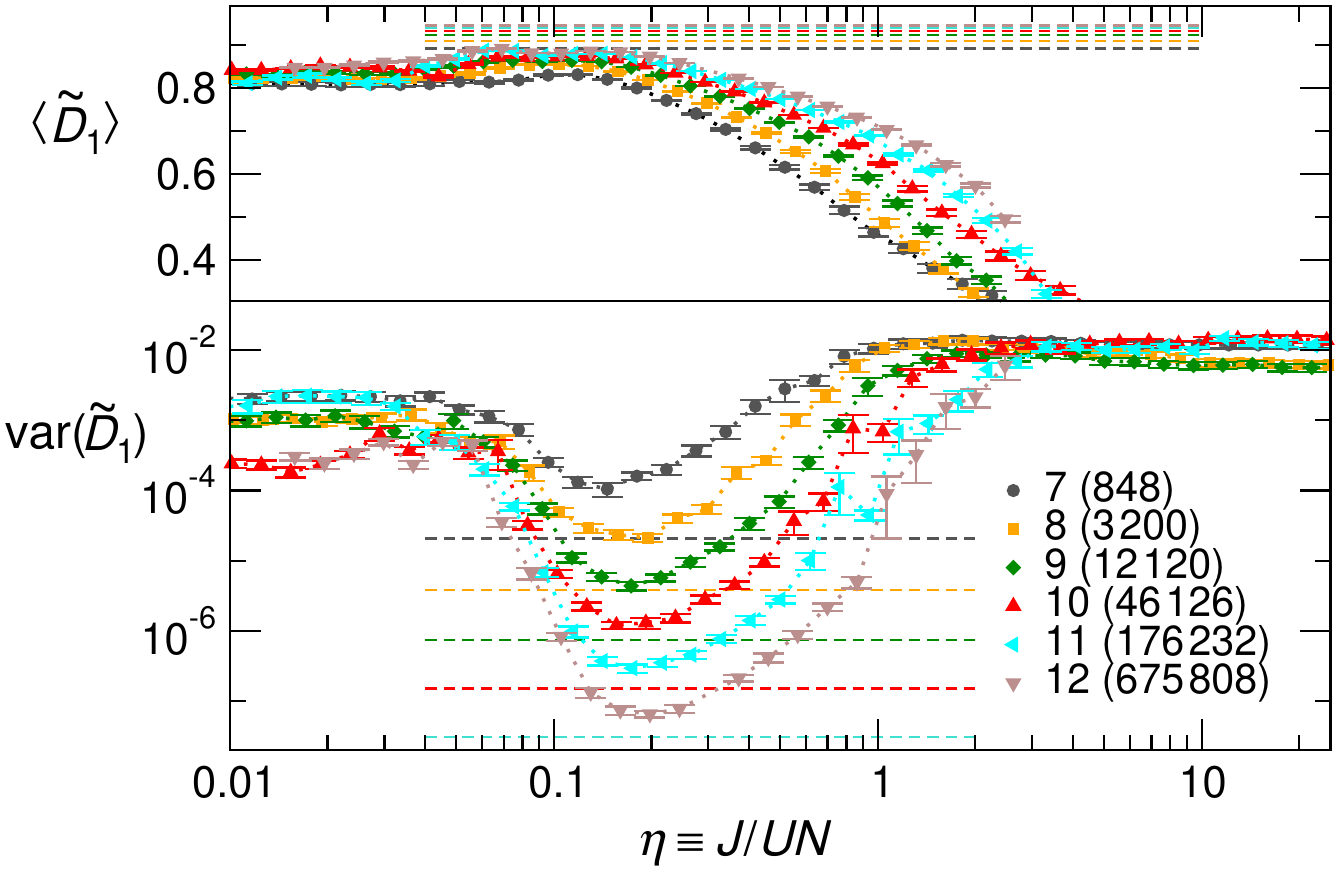}
	\caption{Evolution of $\langle \tilde{D}_1\rangle$ (top) and $\var(\tilde{D}_1)$ (bottom) at $\epsilon=0.5$ versus $\eta$, for varying values of $L$ and size $(\mathcal{N})$ (as indicated by the legend) of the  subspace spanned by the $\pi=-1$ eigenstates of $H_\textrm{tun}$ with HWBC.
		Each data point results from the analysis of the $100$ BHH eigenvectors closest to $\epsilon=0.5$.
		Corresponding GOE values are indicated by dashed lines.}
	\label{fig:D1vsJUN_tun}
\end{figure}

The energy-resolved correlation between the spectral statistics and the eigenvector structure in the eigenbasis of $H_\mathrm{tun}$, as function of rescaled energy $\epsilon$ and hopping strength $\eta$, is presented for $\langle \tilde{D}_1\rangle$ in Fig.~\ref{fig:en_dens_TUN}, and should be compared against Fig.~\ref{fig:en_dens_INT} in the manuscript. \red{To illustrate similarities and differences between the two bases, the color scales were chosen exactly identical in both figures.} Note that the overall evolution of the values of the GFD is inverted as compared to the analysis in the $H_\mathrm{int}$ basis, since the eigenvectors must be highly localized in the limit $\eta\to\infty$.
The evolution of $\var(\tilde{D}_q)$ in the $H_\mathrm{tun}$ eigenbasis shows a region of drastically suppressed GFD fluctuations 
that agrees 
with the corresponding area observed in Fig.~\ref{fig:en_dens_INT}.
The inspection of $\langle \tilde{D}_1\rangle$ at $\epsilon=0.5$, shown in Fig.~\ref{fig:D1vsJUN_tun}, does not reveal the clear formation of a plateau, and the typical GFD values are rather far from the RMT prediction, as also illustrated in Fig.~\ref{fig:ScalingDq} of the manuscript. Despite the basis dependence of the typical GFD, their fluctuations might be a basis-independent figure of merit to identify the emergence of a chaotic regime.

\section{Generalized fractal dimensions for GOE eigenvectors}

\begin{figure}
	\centering
	\includegraphics[width=\columnwidth]{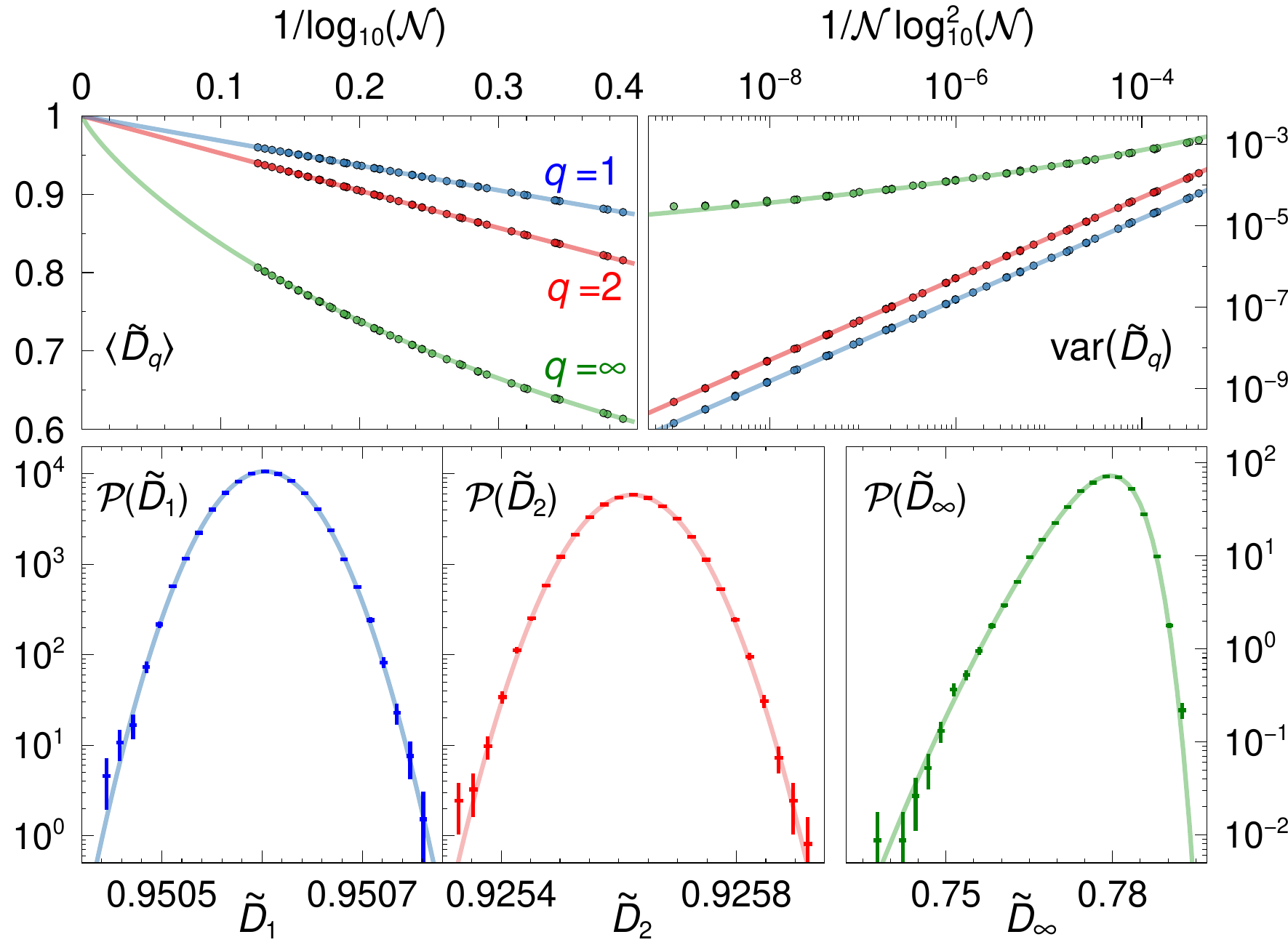}
	\caption{Comparison between analytical and numerical results for the GFD of GOE eigenvectors. Average and variance of $\tilde{D}_q$ as a function of the vector length $\mathcal{N}$ are shown in the upper panels, while the lower plots display probability density functions of $\tilde{D}_q$ for $\mathcal{N}=2\,599\,688$. Solid lines follow from the evaluation of Eqs.~\eqref{eq:D1GOEana}-\eqref{eq:DinfGOEana}; symbols and error crosses (indicating $\pm 1\sigma$) are obtained from the numerical sampling of GOE eigenvectors [$10^4$ for $\langle\tilde{D}_q\rangle$, $\var(\tilde{D}_q)$, and  $5\times10^4$ for the distributions].}
	\label{fig:GOEnumerics}
\end{figure}

Since $\tilde{D}_1=-(\ln\mathcal{N})^{-1}\sum_\alpha |\psi_\alpha|^2\log_\mathcal{N}|\psi_\alpha|^2$, the calculation of its mean and variance can be carried out exactly from the known one- and two-intensity distributions of GOE eigenvectors \cite{Haake2004}. For $q=2$, the analytical calculations for the typical GFD are rather challenging, but the results for the ensemble-averaged GFD (obtained from the arithmetic average of the $R_q$ moments, i.e., $\langle\tilde{D}_2^\mathrm{(ens)}\rangle=- \log_\mathcal{N} \langle R_2\rangle$) provide excellent approximations \cite{Backer2019},
\begin{align}
\langle \tilde{D}_1 \rangle_\mathrm{GOE} &= \frac{H_{\NN/2}-2+\ln 4}{\ln \NN}, \label{eq:D1GOEana}\\
\langle \tilde{D}_2^\mathrm{(ens)} \rangle_\mathrm{GOE} &= \frac{\ln(\mathcal{N}+2)-\ln3}{\ln\mathcal{N}},
\end{align}
where $H_{n}=\sum_{k=1}^n \frac{1}{k}$ is the harmonic number, and
\begin{align}
\var(\tilde{D}_1) &=  \frac{(3\pi^2-24)(\NN+2)-8}{2(\NN+2)^2 \ln ^2\NN} -\frac{\psi^{(1)}(2+\NN/2)}{\ln ^2\NN}, \\
\var(\tilde{D}_2^\mathrm{(ens)}) &= \frac{8 (\mathcal{N}-1)}{3 (\mathcal{N}+4) (\mathcal{N}+6) \ln ^2(\mathcal{N})}, \label{eq:D2GOEvar}
\end{align}
where $\psi^{(1)}$ denotes the first derivative of the digamma function (see Eq.~5.2.2 in Ref.~\cite{NIST:DLMF}).
We emphasize the importance of using the two-intensity distribution to calculate the variance of $\tilde{D}_1$. The correlation among the intensities induced by normalization plays a crucial role in obtaining  the correct result. The validity of Eqs.~\eqref{eq:D1GOEana}-\eqref{eq:D2GOEvar} is borne out by the numerical data (cf.~Fig.~\ref{fig:GOEnumerics}).

We found that the probability density functions of $\tilde{D}_1$ and $\tilde{D}_2$ are very well described by Gaussians with the corresponding mean and variance given above, as demonstrated in Fig.~\ref{fig:GOEnumerics}.

For $q=\infty$, we deal with the extreme statistics of the eigenvector intensities. In this case, one can proceed in the way suggested in Ref.~\cite{Lakshminarayan2008}. In order to find the distribution for the variable $t\equiv \max_\alpha|\psi_\alpha|^2$, we neglect the correlation among the intensities induced by the eigenstate normalization (such correlation does not seem to be crucial in this case).
From the Porter-Thomas distribution $P(\left|\psi_\alpha\right|^2)$ for the wavefunction intensities for large $\NN$ \cite{Haake2004}
the cumulative distribution function of $t$ thus reads $F(t,\mathcal{N})=\prod_{\alpha=1}^\NN \int_0^t \;\mathrm{d}\left|\psi_\alpha\right|^2\;  P(\left|\psi_\alpha\right|^2)=
\left[\Erf(\sqrt{t\mathcal{N}/2})\right]^\mathcal{N}$, and consequently its PDF can be written as
\begin{equation}
\rho(t,\mathcal{N})=\frac{\mathcal{N}^{3/2}}{\sqrt{2\pi t}}e^{-t\mathcal{N}/2}\left[\Erf(\sqrt{t\mathcal{N}/2})\right]^{\mathcal{N}-1}.
\end{equation}
The PDF for $\tilde{D}_\infty=-\log_\mathcal{N}t$ is thus given by
\begin{equation}
\mathcal{P}(\tilde{D}_\infty)=\rho (\mathcal{N}^{-\tilde{D}_\infty},\mathcal{N})\,\mathcal{N}^{-\tilde{D}_\infty}\ln\mathcal{N},
\end{equation}
which is in excellent agreement with the numerics, as we demonstrate in Fig.~\ref{fig:GOEnumerics}.
After doing an appropriate change of variable and integrating by parts (neglecting one term that decreases exponentially with $\mathcal{N}$), one can estimate the moments of $\tilde{D}_\infty$ from
\begin{equation}
\langle\tilde{D}_\infty^k\rangle=\frac{(-1)^{k-1}2k}{\ln^k\mathcal{N}}\int_{1/\sqrt{2}}^{\sqrt{\mathcal{N}/2}} dx \frac{[\Erf(x)]^\mathcal{N}}{x}\ln^{k-1}(2x^2/\mathcal{N}).
\label{eq:Dinfmoments}
\end{equation}
After an adequate treatment of $[\Erf(x)]^\mathcal{N}$ for large $\mathcal{N}$, we find for $k=1$
\begin{align}
\langle \tilde{D}_\infty\rangle_\mathrm{GOE} =& 1-\frac{\ln (2\ln\mathcal{N})}{\ln\mathcal{N}} +\frac{\ln(\ln^2(2)\pi\ln\mathcal{N})}{2\ln^2\mathcal{N}} \notag \\
&+O\left(\ln^2\ln\mathcal{N}/\ln^3\mathcal{N}\right),
\label{eq:DinfGOEana}
\end{align}
which provides a remarkable description of the numerical data, as shown in Fig.~\ref{fig:GOEnumerics}.
It is worth noting that the leading correction for $\tilde{D}_\infty$ exhibits the generic dependence expected for the extreme statistics of multifractal eigenvectors \cite{Fyodorov2015}.
The calculation of the leading terms of the second moment, and thus of the variance, proves to be more involved, although analytical inspection suggests that $\var(\tilde{D}_\infty)_\mathrm{GOE}\sim\ln^{-4}\mathcal{N}$ as $\mathcal{N}\to\infty$. In any case, $\langle\tilde{D}_\infty^2\rangle$ can be estimated by evaluating numerically Eq.~\eqref{eq:Dinfmoments}.
\section{Departure of $\boldsymbol{\mathcal{P}(\tilde D_q)}$ from GOE and finite-size corrections to $\boldsymbol{\langle \tilde D_q\rangle}$}

\begin{figure}
	\centering
	\includegraphics[width=.9\columnwidth]{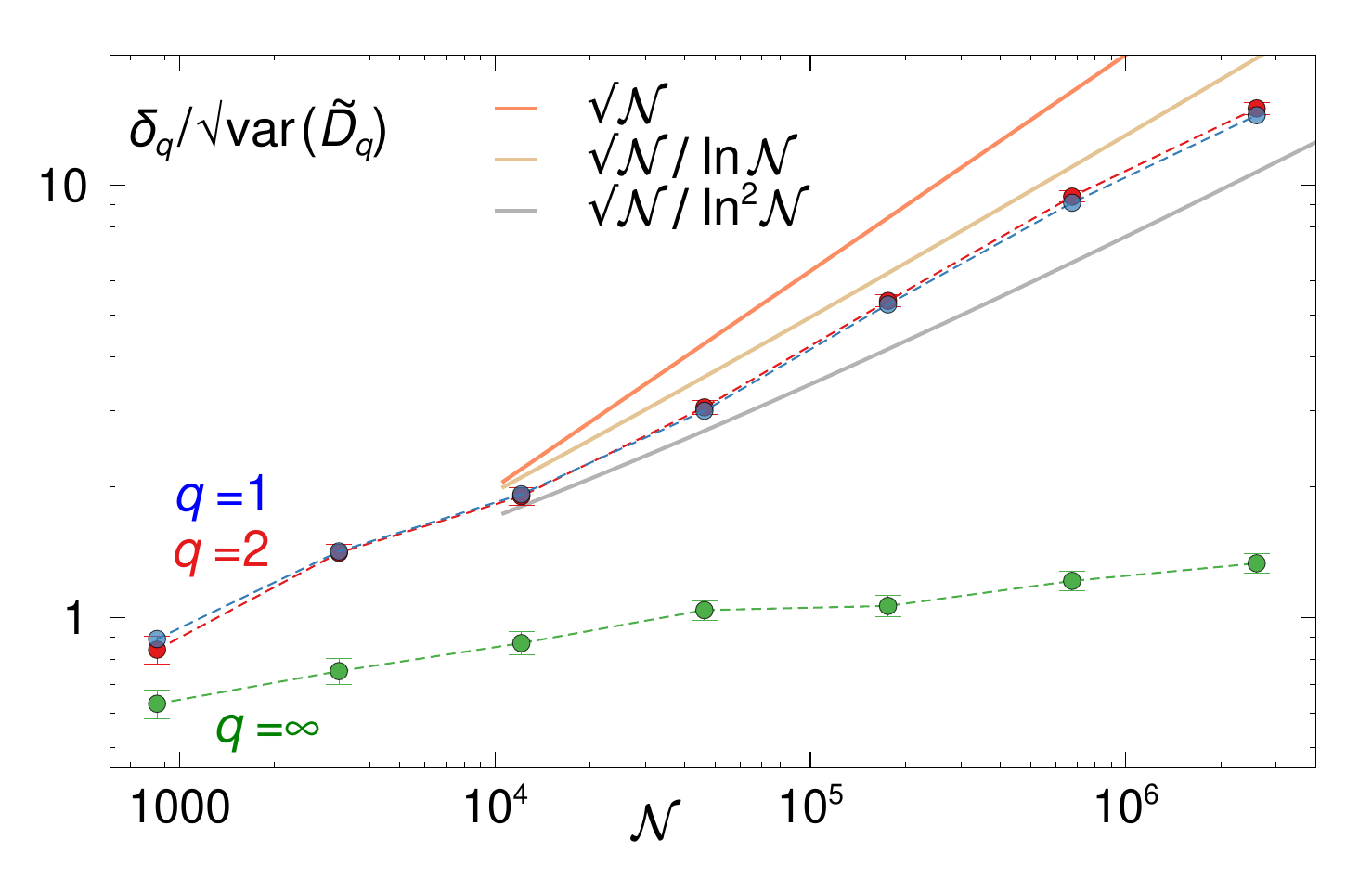}
	\caption{\red{Distance $\delta_q/\sqrt{\var(\tilde D_q)}$ between the distributions of $\tilde{D}_q$ for the BHH and GOE eigenvectors as a function of $\mathcal{N}$, as shown in panel (c) of Fig.~fig:PDFs} in the manuscript. Solid lines indicate different functional dependencies of the form $\sqrt{\mathcal{N}}/\ln^{n-1}\mathcal{N}$, for $n=1,2,3$.}
	\label{fig:PDFdistance}
\end{figure}

\red{
	The behavior of the distance between the GFD distributions for BHH and GOE eigenvectors, evaluated via $d_q(\mathcal{N})\equiv\delta_q/\sqrt{\var(\tilde{D}_q)}$, reveals how 
	the finite-$\mathcal{N}$ corrections in $\left\langle\tilde{D}_q\right\rangle$ and $\left\langle\tilde{D}_q\right\rangle_\mrm{GOE}$ differ from each other. 
	For $q=1,2$, the numerical data shows that the leading correction to both $\left\langle\tilde{D}_q\right\rangle$ and $\left\langle\tilde{D}_q\right\rangle_\mrm{GOE}$ is $\sim 1/\ln\NN$. If the coefficient of this term differs slightly between BHH and GOE, it ensues that $\delta_q\equiv\left\langle\tilde{D}_q\right\rangle_\mrm{GOE}-\left\langle\tilde{D}_q\right\rangle\sim 1/\ln\NN$. Since  $\sqrt{\var(\tilde{D}_q)}\sim 1/\sqrt{\NN}\ln\NN$, this would lead to
	\begin{align}
	d_q(\mathcal{N})\sim \sqrt{\NN}.
	\end{align} 
	On the other hand, if the leading term coincides for both models, 
	then $\delta_q$ will scale as the dominant subleading term in either $\left\langle\tilde{D}_q\right\rangle$ or $\left\langle\tilde{D}_q\right\rangle_\mrm{GOE}$. 
	Assuming subleading corrections in the BHH of the form $\ln^{-n} \NN$, $n\geq 2$, one obtains 
	\begin{align}
	d_q(\mathcal{N})\sim \frac{\sqrt{\NN}}{\ln^{n-1} \NN}.
	\end{align}
	Alternatively, if the first subleading term to $\left\langle \tilde{D}_q\right\rangle$ had the same functional dependence on $\NN$ as for GOE, one would find $\delta_q\sim 1/\NN\ln \NN$ and hence
	\begin{align}
	d_q(\mathcal{N}) \sim \frac{1}{\sqrt{\NN}},
	\end{align}
	which would decrease with $\NN$, in contrast to the numerically found behavior. 
	We can therefore rule out this latter possibility. 
}

\red{In Fig.~\ref{fig:PDFdistance}, we compare $d_{1,2}(\mathcal{N})$ with different scalings of the form $\sqrt{\NN}/\ln^{n-1} \NN$, for $n=1,2,3$. The tendency of the data for large $\NN$ seems to be best described by the case $n=2$, 
	which implies that most likely the dominant finite-size correction to $\left\langle\tilde{D}_{1,2}\right\rangle$ is exactly described by GOE while the next term is of the form $1/\ln^2\NN$. 
}

\end{document}